\RequirePackage{fix-cm}
\documentclass[twocolumn,epjc3]{svjour3}  
\smartqed  
\usepackage{amsmath}
\RequirePackage{graphicx}
\RequirePackage{mathptmx}      
\usepackage[colorlinks,citecolor=blue,urlcolor=blue,linkcolor=blue]{hyperref}

\journalname{Eur. Phys. J. C}

\begin{document}

\title{Early dark energy induced by non-linear electrodynamics}
\subtitle{Early dark energy as non-linear electrodynamics}

\author{H. B. Benaoum\thanksref{e1,addr1}
        \and
        Luz Ángela García\thanksref{e2,addr2}
        \and
        Leonardo Castañeda\thanksref{e3,addr3}
}
\thankstext{e1}{e-mail: hbenaoum@sharjah.ac.ae}
\thankstext{e2}{e-mail: lgarciap@ecci.edu.co}
\thankstext{e3}{e-mail: lcastanedac@unal.edu.co}

\institute{Department of Applied Physics and Astronomy, University of Sharjah, United Arab Emirates \label{addr1}
\and Universidad ECCI, Cra. 19 No. 49-20, Bogot\'a, Colombia, C\'odigo Postal 111311 \label{addr2}
\and Observatorio Astronómico Nacional, Universidad Nacional de Colombia \label{addr3}
}

\date{Received: date / Accepted: date}

\maketitle

\begin{abstract}
In this work, we introduce a parametrization of early dark energy that mimics radiation at early times and governs the present acceleration of the Universe. We show that such parametrization models non-linear electrodynamics in the early Universe and investigate the cosmological viability of the model. In our scenario, the early dark energy is encoded in the non-linearity of the electromagnetic fields through a parameter $\beta$ that changes the Lagrangian of the system, and the parameters $\gamma_s$ and $\alpha$, that define the departure from the standard model constant equation of state. We use a Bayesian method and the modular software \textsc{CosmoSIS} to find the best values for the model's free parameters with precomputed likelihoods from Planck 2018, primordial nucleosynthesis data, inferred distances from different wide galaxy surveys and luminosity distances of SNIa from Pantheon and SH0ES, such that $\gamma_s =$ 0.468 $\pm$ 0.026 and $\alpha =$ -0.947 $\pm$ 0.032, as opposed to $\Lambda$CDM where $\gamma_s = \beta =$ 0 and there is no equivalence for the $\alpha$ parameter. Our results predict an earlier formation of the structure and a shorter age of the Universe compared with the canonical cosmological model. One of the main findings of our work is that this kind of dark energy alleviates the ongoing tensions in cosmology, the Hubble tension and the so-called $\sigma_8$ tension, which predicted values by our model are H$_o =$ 70.2 $\pm$ 0.9 km/s/Mpc and $\sigma_8 =$ 0.798 $\pm$ 0.007. The reported values lie between the inferred values inferred from early and late (local) Universe observations. Future observations will shed light on the nature of the dark energy, its impact on the structure formation, and its dynamics.
\end{abstract}

\section{Introduction}
\label{intro}
Despite its simplicity, the standard $\Lambda$CDM model has successfully explained many cosmological observations \cite{caldwell2003phantom,ade2016planck}. However, with the improvement of the observational data, significant tensions between $\Lambda$CDM and some data might indicate new physics beyond the $\Lambda$CDM concordance model. \newline

Based on canonical cosmic model, the observed present value of the Hubble parameter is H$_o =$ 67.4 $\pm$ 0.5 km/s/Mpc 68\% CL \cite{planck2020} from the Planck 2018 Cosmic Microwave Back\-ground (CMB) which has $5 \sigma$ tension discrepancy with a model-independent local measurement H$_o =$ 73.04 $\pm$ 1.04 km/s/Mpc at 68\% CL \cite{riess2022} from the Hubble Space Telescope (HST) observations of variable Cepheids. In addition to the observed present value of the Hubble parameter H$_o$ disagreement, a particular important tension, related to the amplitude of the density fluctuations quantified in terms of the parameter $S_8$, between CMB data and the cosmic shear and redshift space distortions (RSD). A lower $S_8$ value ranging between $0.703$ and $0.782$ have been inferred from redshift surveys data and weak lensing measurements compared to the one estimated from Planck data, $S_8 =0.834 \pm 0.016$ \cite{planck2020}. From ACT+WMAP analysis, $S_8 = 0.840 \pm 0.30$ \cite{aiola2020atacama}, assuming $\Lambda$CDM model. Although the $S_8$ tension could be related to systematic errors, it might hint at a possible new physics beyond the standard $\Lambda$CDM model. Many possible alternative new physics solutions have been proposed to solve these tensions, which require modifications in the early Universe into the pre-re\-com\-bi\-na\-tion era and the late history of the Universe. \newline

Early dark energy models have caught great interest in the community in the past few years. Early-time solutions aim to suppress the value of the sound horizon by injecting a fraction of energy before recombination without spoiling the fit to the CMB and Baryon Acoustic Oscillations (BAO) data. In particular, if 10\% of the total energy density is allowed in the Hubble parameter during the radiation era (3500 $< z <$ 5000), the Hubble parameter prediction from the early and late Universe significantly drops \cite{smith2021,klypin2021}. Thus, as alternatives to the cosmological constant $\Lambda$, early dark energy models are a compelling method to treat the ongoing tensions and introduce dynamics in stages when dark energy has been assumed to not play a role in the Universe. \newline
There are currently different models to describe early dark energy, among them: \cite{garcia2011} introduce additional degrees of freedom in Hubble parameter through sterile neutrinos; a modified Chaplygin gas that simultaneously describes the dynamics of dark matter and dark energy \cite{bento2002}; effective pa\-ra\-me\-te\-ri\-za\-tions that evolve from a non-negligible contribution during the radiation-domination epoch to the current accelerated expansion of the Universe \cite{garcia2021}. An extensive summary of models that compete with the cosmological constant $\Lambda$ effect in the Universe´s current expansion can be found in \cite{schoneberg2021,poulin2023}. \newline
Another alternative to model dark energy is non-linear electrodynamics, replacing the original Maxwell Lagrangian with a non-linear electrodynamics (NLED) La\-gran\-gian. 
Non-li\-near electrodynamics \cite{born1934foundations} is a generalization of Max\-well's electromagnetic theory and, when coupled to gravity, produces a negative pressure that tends to accelerate the expansion of the Universe at early and/or late stages. In recent years, non-linear electrodynamics has been the object of a significant amount of interest in cosmology and astrophysics \cite{vollick2003anisotropic,moniz2002quintessence,garcia2000born,camara2004nonsingular,elizalde2003born,novello2004nonlinear,kruglov2015universe,ovgun2018falsifying,ovgun2017inflation,benaoum2021}. \newline

The outline of the paper is as follows. In section~\ref{section2}, we propose a more general parametrization of an early dark energy model that mimics radiation at early times and has an accelerated expansion at late times. Such parametrization incorporates a possible transition of the equation of state during the evolution of the Universe. In section~\ref{section2non}, we present our phenomenological model based on non-linear electrodynamics that links the non-linearity of the electromagnetic fields to the early dark energy. To our knowledge, this work is the first to address the generation of early dark energy from the non-linearity of electromagnetic fields. In section~\ref{section3}, we extensively discuss the method to find the best values for the model's free parameters and present the results with Bayesian statistical inference. Section~\ref{section4} shows some cosmological tests we submit our model with the best-fits calculated in the previous section. Finally, we summarize our findings and conclusions in section~\ref{section5}.

\section{Parametrization of Early Dark Energy}
\label{section2}
We propose a novel parametrization of dark energy. It is particularly suited to describe a small but non-negligible amount of dark energy at early stages and an accelerated expansion in the late phases of the Universe. In this section, we use the modified Chaplygin gas model \cite{benaoum2002,benaoum2012,benaoum2019extensions} approach to parametrize the dark energy and investigate the effect of adding  a small fraction of dark energy in the Universe's early evolution. \newline

The Chaplygin gas was first introduced by S. Chaplygin \cite{chaplygin1904} at the beginning of the 20th century to describe the lifting force experienced by the winds of an airplane when it is in the air. The gas creates a negative pressure, generating an effective opposite effect to the gravitational force.\newline
The idea is brought back in the early 2000s with the advent of dark energy by \cite{benaoum2002}. The author presented a model that unifies dark matter and energy in a generalized Chaplygin gas. Later on, \cite{benaoum2012} presented a modified Chaplygin gas that is particularly suited to describe a small but non-negligible amount of dark energy at early stages and an accelerated expansion at late phases of the Universe. 
Our model is characterized by three parameters: the present-day value of dark energy density $\rho_{de,o} \simeq 1.7 \times 10^{-119} M_p^4$ where $M_p = 2.435 \times 10^{18} ~$GeV is the reduced Planck mass, the transition scale factor and a third parameter that controls the rapidity of this transition. In this scenario, the dark energy mimics radiation in the early Universe and accelerates its expansion at late times. \newline

To motivate our investigation, we consider a spatially flat Friedmann-Lemaitre-Robertson-Walker (FLRW) geometry described by the line element,
\begin{eqnarray}
d s^2 & = & - dt^2 + a^2 (t) \left(dx^2 + dy^2 + dz^2 \right),
\end{eqnarray} 
\noindent where $a (t)$ is the expansion scale factor. The expansion dynamics are governed by Friedmann's equations,
\begin{eqnarray} 
    H^2 & = & \frac{8 \pi G}{3} ~\rho, \nonumber \\
\dot{H} & = & - 4 \pi G \left( \rho + p \right).
\end{eqnarray}
Here $H = \dot{a}/a$ is the Hubble rate parameter, $\rho$ and $p$ are the total energy and pressure of the fluid species filling the Universe. The total cosmic fluid obeys the continuity equation,
\begin{eqnarray}
\dot{\rho} + 3 H \left( \rho + p \right) & = & 0.
\end{eqnarray}

Throughout the present work, we assume that the Universe is composed of radiation, pressure-less dark matter, and dark energy. The total energy density and the total pressure are just the sum of the contributions of all species,
\begin{eqnarray} 
\rho &  = & \rho_r + \rho_m + \rho_{de}, \nonumber\\
p & = & p_r + p_{de}, 
\end{eqnarray}
where the sub-index $r,m$ and $de$ stands for radiation, matter and dark energy. The energy densities of the radiation and matter are given by:
\begin{eqnarray}
\rho_r & = & \rho_{r,o} ~a^{-4} \nonumber \\
\rho_m & = & \rho_{m,o} ~a^{-3} 
\end{eqnarray}
The dark energy is assumed to evolve independently, and its energy density is expressed by the continuity equation:
\begin{eqnarray}
\dot{\rho}_{de} + 3 H \left(\rho_{de} + p_{de} \right) & = & 0. 
\end{eqnarray}
From this, it is straightforward to determine the dark energy equation of state (EoS) $\omega_{de} = \frac{p_{de}}{\rho_{de}}$ as:
\begin{eqnarray}
\omega_{de} & = & -1 - \frac{a}{3} \frac{d \ln \rho_{de}}{d a}. 
\label{eos}
\end{eqnarray}
It is possible to express the early dark energy $\rho_{de} (a)$ as:
\begin{eqnarray}
\rho_{de} (a) & = & \rho_{de,o} ~G (a)^\frac{1}{1+\alpha} 
\label{rho}
\end{eqnarray}
\noindent with $\alpha > -1$. The term   $\rho_{de,0}$ is the energy density at present time (i.e., $a=1$) and  $G (a)$ is a generic function which we parametrize as having the following functional form,
\begin{eqnarray}
G (a)  & = & 1 - \gamma_s + \gamma_s ~a^{- 4 (1+\alpha)}.
\label{param}
\end{eqnarray}
\noindent Note that if $\gamma_s =$ 0, $G (a) =$ 1, and the model reduces to the standard $\Lambda$CDM model. This condition defines the lower limit for $\gamma_s$ prior.\newline

The solution to the dark energy EoS \eqref{eos} can be written in terms of the scale factor as:
\begin{eqnarray}
\omega_{de} & = & -1 + \frac{4}{3} \frac{1}{1+ \frac{1-\gamma_s}{\gamma_s} a^{4 (1+ \alpha)}} 
\label{eos1}
\end{eqnarray}
Equations~\eqref{rho}, \eqref{param} and \eqref{eos1} yield several qualitative and interesting insights into the behavior of $\rho_{de}$ and $\omega_{de}$\footnote{It is worth noting that the parametrization we present in this document is motivated in a completely different scenario than the model presented in \cite{garcia2021}. However, an unmistakable resemblance exists between \eqref{eos1} and the early dark energy model described by \cite{garcia2021}. The close similarity in the functional forms of both parametrizations is explained by the fact that both models assume that the equation of state transitions between the value for radiation in the early Universe to $-$1 in the late epochs when dark energy is dominant.}. First, note that in our parametrization, the EoS of dark energy converges to $\omega_{de} \rightarrow -1$ at late times, which implies that dark energy is the dominant component. Also, the condition $\alpha > -$ 1 must be fulfilled to have the usual cosmic domination eras in place. The energy density must be subdominant at BBN/CMB scale to satisfy the constraints $\rho_{de}/\rho_r \leq 0.086$ \cite{artymowski2021emergent}. Thus one should expect that the ratio $\rho_{de}/\rho_r$ at radiation era to be,
\begin{eqnarray}
\frac{\rho_{de}}{\rho_r}\Big|_{rad} \simeq \frac{\rho_{de,o}}{\rho_{r,o}} \gamma_s^{\frac{1}{\alpha+1}} < 1
\end{eqnarray}
One can see that this ratio depends on the parameters $\gamma_s$ and $\alpha$ as $\rho_{de,o}$ and $\rho_{r,o}$ are fixed by the cosmological data.\newline

\begin{figure}[hbtp]
\centering
  \includegraphics[width=.5\textwidth]{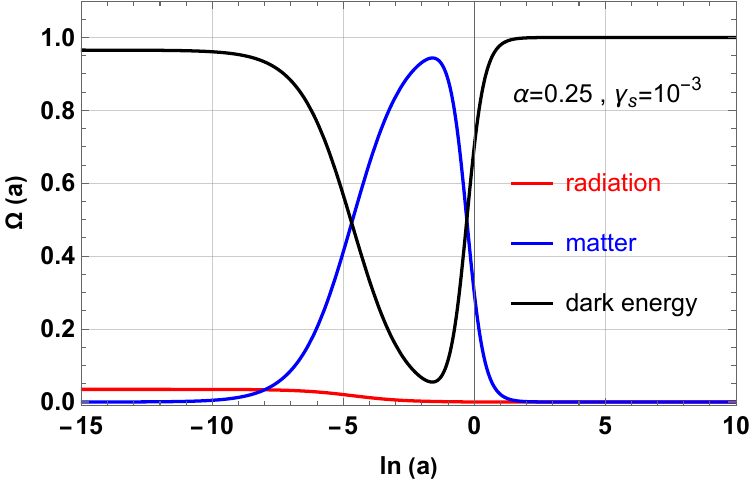} 
\caption{Evolution of the energy densities $\Omega (a)$ for radiation, matter and dark energy for 
$\gamma_s =10^{-3}$ and $\alpha =0.25$. Here $\Omega_{r,o} =10^{-4}$, $\Omega_{m,o}=0.3$ and $\Omega_{de,o}=1-\Omega_{m,o}- \Omega_{r,o}$.}
\label{fig:1}
\end{figure}

\begin{figure}[hbtp]
\centering
  \includegraphics[width=.5\textwidth]{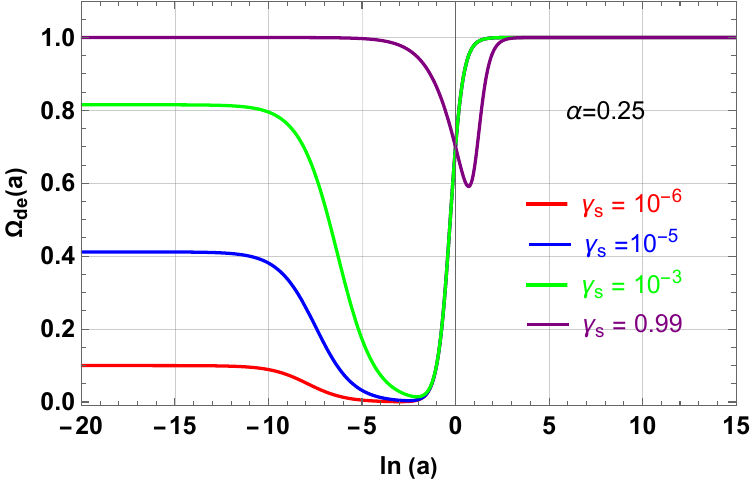} 
\caption{Evolution of the energy density for dark energy $\Omega_{de} (a)$ for $\alpha =0.25$ and different values of $\gamma_s$. Here $\Omega_{r,o} =10^{-4}$, $\Omega_{m,o}=0.3$ and $\Omega_{de,o}=1-\Omega_{m,o}- \Omega_{r,o}$.}
\label{fig:defraction_diff_al}
\end{figure}

\begin{figure}[hbtp]
\centering
  \includegraphics[width=.5\textwidth]{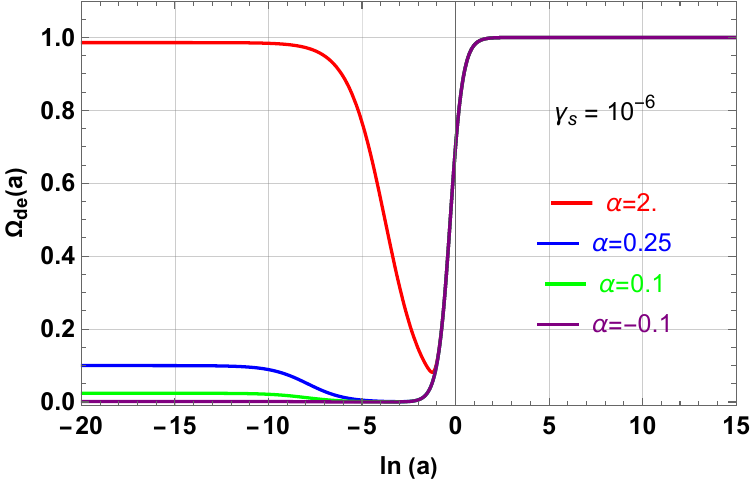} 
\caption{Evolution of the energy density for dark energy $\Omega_{de} (a)$ for $\gamma_s =$ 10$^{-6}$ and different values of $\alpha$. Here $\Omega_{r,o} =10^{-4}$, $\Omega_{m,o}=0.3$ and $\Omega_{de,o}=1-\Omega_{m,o}- \Omega_{r,o}$.}
\label{fig:3}
\end{figure}

\begin{figure}[hbtp]
\centering
  \includegraphics[width=.5\textwidth]{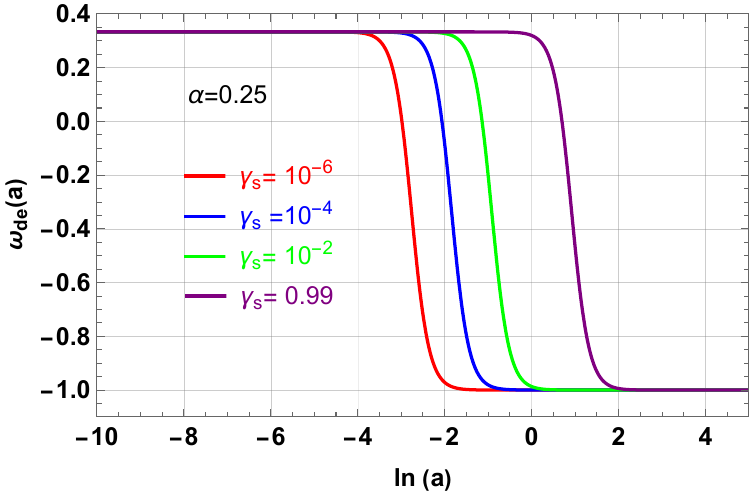}
\caption{Evolution of the equation of state of dark energy $\omega_{de} (a)$ for $\alpha =0.25$ and different values of $\gamma_s$. Here $\Omega_{r,o} =10^{-4}$, $\Omega_{m,o}=0.3$ and $\Omega_{de,o}=1- \Omega_{m,o} - \Omega_{r,o}$.}
\label{fig:eos_diff_as}
\end{figure}

In Figures~\ref{fig:1} to \ref{fig:eos_diff_as}, we show the exact behavior of $\Omega_{de}=\rho_{de}/\rho_{cr}$ and $\omega_{de}$ as a function of the scale factor $a$ for different values of $\alpha$ and $\gamma_s$. An example of the evolution of the density parameter $\Omega_{de}$ for $\alpha = 0.25$ and $\gamma_s = 10^{-3}$ is shown in Figure~\ref{fig:1}. In Figures~\ref{fig:defraction_diff_al} and \ref{fig:3}, we show the behavior of $\Omega_{de}$ by fixing one of the parameters and varying the other one. One can see that the evolution of $\Omega$ is extremely sensitive to the values of $\gamma_s$ and $\alpha$.\newline

We highlight that there is a scale factor $a_t$ in the matter domination era that determines the transition of the gas from the radiation epoch to the current expansion, given by:
\begin{eqnarray}\label{at}
a_t & = & \left( \frac{\gamma_s}{1- \gamma_s}  \right)^{\frac{1}{4 (1+ \alpha)}}
\end{eqnarray}
The way in which $\rho_{de}$ and $\omega_{de}$ evolve with the scale factor is extremely sensitive to $A_s$ and $\alpha$ where the transition scale factor $a_t$ controls the time of transition and $\alpha$ defines its duration so that a small value for $\alpha$ corresponds to a shorter transition period. The dark energy equation of state $\omega_{de}$ follows peculiar behaviours, starting with $\omega_{de}= \frac{1}{3}$ for $a << a_t$ during radiation era, passing by  $\omega_{de}=0$ for $a= \left(\frac{\gamma_s}{3 (1 - \gamma_s)} \right)^{\frac{1}{4 (1+ \alpha)}}$ and $\omega_{de} = -1$ in recent epochs (see Figure~\ref{fig:eos_diff_as}).\newline

One of the important ways to check the causality of the universe to persists is the adiabatic squared speed of the sound, which is given by:

\begin{eqnarray}
c_s^2 & = & \frac{\delta p_{de}}{ \delta \rho_{de}} =  \omega_{de} - \frac{\omega^{\prime}_{de}}{3 (1+ \omega_{de})} 
\nonumber \\
& = & \frac{1}{3} + \frac{4 \alpha}{3} \frac{1}{1 + \left(\frac{a_t}{a} \right)^{4 (1+ \alpha)}} ~~.
\label{speed}
\end{eqnarray}
\noindent where primes denotes derivatives with respect to $\ln (a)$.\newline 

\begin{figure}[hbtp]
\centering
  \includegraphics[width=.5\textwidth]{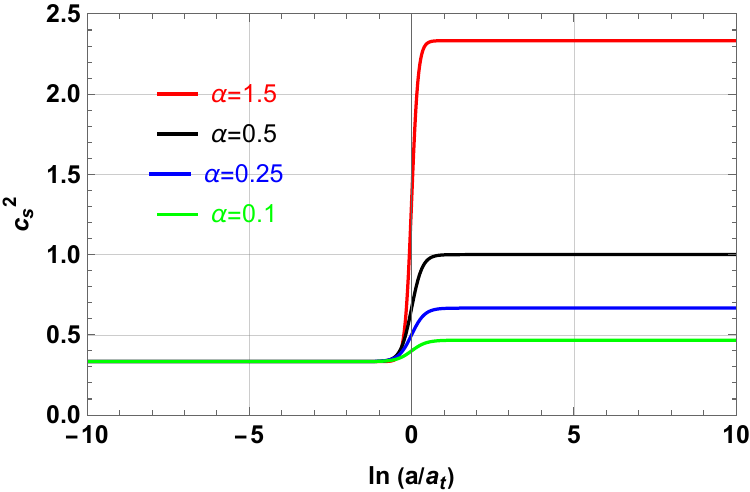} 
\caption{Variation of the speed of sound $c_s^2$ as a function of $\ln (a/a_t)$ for different values of $\alpha$.}
\label{fig:cs2}
\end{figure}

For the causality condition, the speed of the sound must be less than the local light speed $c_s^2 \leq 1$. A positive value of $c_s^2$ represents a stable model, whereas a negative value of $c_s^2$ indicates the instability of the model. It follows from our model that $c_s^2 \approx \frac{1}{3}$ for $a << a_t$ and $c_s^2 \approx \frac{1}{3} (1 + 4 \alpha)$ for $a >> a_t$ which shows that the causality and classical stability is satisfied for $\alpha <0.5$.
In Figure \ref{fig:cs2}, we plot the behavior of the square speed of sound as a function of the number of $\ln(a/a_t)$ for different values of $\alpha$. As shown in the figure, one sees clearly for which values of $\alpha$ the causality and classical stability are satisfied (i.e., $\alpha >$ 0.5 breaks the model stability). \newline

Finally, we define the dimensionless Hubble parameter $E(z)$ as function of the redshift $z$, as follows: 
\begin{align}\label{e_z}
E (z)^2 & =\frac{H (z)^2}{H_0^2}  \nonumber\\
& = \Omega_{o,r} (1+z)^4 + \left(\Omega_{o,b} +\Omega_{o,c} \right) (1+z)^3 \nonumber\\
& +\Omega_{de,o} \left(1 - \gamma_s + \gamma_s (1+z)^{4 (1+\alpha)} \right)^\frac{1}{1+\alpha},
\end{align} 
where $\Omega_{o,r}$, $\Omega_{o,b}$, $\Omega_{c,o}$, $\Omega_{de,o}=1- \Omega_{o,r}-\Omega_{o,b}-\Omega_{c,o}$ are the radiation, baryons, cold dark matter, and dark energy  density fractions at present time, respectively. Also, we have assumed here the relation between the scale factor $a$ and the redshift $z$, given by the equality $a = \frac{1}{1+z}$.

\section{Non-linear electrodynamics as an early dark energy}\label{section2non}
Here we discuss one possible avenue to describe the nature of the early dark energy. Non-linear electrodynamics is expected to play a crucial role in the evolution of the Universe. For this purpose, we propose the following non-linear electrodynamics Lagrangian \cite{benaoum2021,benaoum2023inflation},
\begin{eqnarray}
{\cal L}_{nled} & \equiv & -{\cal F}  ~f \left( {\cal F} \right), 
\end{eqnarray}
where ${\cal F} = \frac{1}{4} F_{\mu \nu} F^{\mu \nu}= \frac{1}{2} (B^2 - E^2)$ and $f \equiv f ({\cal F})$ is a functional that depends on the field strength ${\cal F}$. \\

The energy-momentum tensor for this Lagrangian density ${\cal L}$, is given by: 
\begin{eqnarray}
T^{\mu \nu} & = & H^{\mu \lambda} F^{\nu}_{\lambda} - g^{\mu \nu} {\cal L}_{nled},
\end{eqnarray}
where $H^{\mu \lambda}$ is given by:
\begin{eqnarray}
H^{\mu \lambda} & = & \frac{\partial {\cal L}_{nled}}{\partial F_{\mu \lambda}} = \frac{\partial {\cal L}_{nled}}{\partial {\cal F}} F^{\mu \lambda}.
\end{eqnarray}
For the Lagrangian density of our model, the energy-\-momentum tensor becomes: 
\begin{eqnarray}
T^{\mu \nu} & = & - \left( f + {\cal F} \frac{d f}{{d \cal F}} \right)  F^{\mu \lambda} F^{\nu}_{\lambda} + 
g^{\mu \nu} {\cal F} f, 
\end{eqnarray}
where the energy density $\rho$ and pressure $p$ can be expressed as: 
\begin{eqnarray}
\rho & = & {\cal F} f - E^2 \left( f + {\cal F} \frac{d f}{d {\cal F}}\right), \nonumber \\
p & = & - {\cal F} f + \frac{2 B^2 - E^2}{3} \left(f + {\cal F} \frac{d f}{d {\cal F}}  \right).
\end{eqnarray}
Here we assume that the wavelength of the electromagnetic waves is typically smaller than the space-time curvature. Thus, we can use the spatial average of fields proposed by \cite{tolman1930}, which defines the volumetric spatial average of a quantity $Y$ as,
\begin{eqnarray*}
\langle Y \rangle & = & \lim_{V \to V_0} \int Y \sqrt{-g} ~d^3 x
\end{eqnarray*}
where $g$ denotes the denotes the determinant of the metric tensor, $V = \int \sqrt{-g} ~d^3 x$ and $V_0$ stands for the time dependent large spatial volume.
In this procedure, the means values of the electric and magnetic fields are given by the Tolman relations, as
\begin{eqnarray*}
\langle E_i \rangle = 0, ~~ \text{} \langle B_i \rangle =0,~~\text{ }\langle
E_{i}B_{j}\rangle =0,
\end{eqnarray*}
\begin{eqnarray*}
\langle E_{i}E_{j}\rangle =\frac{1}{3}E^{2}g_{ij},\text{ }\langle
B_{i}B_{j}\rangle =\frac{1}{3}B^{2}g_{ij}.
\end{eqnarray*}
The latter procedure does not break the isotropy of the Fried\-mann-Roberson-Walker metric. In what follows, the brackets $\langle ~~ \rangle$ will be omitted for simplicity. \newline

In the present work, we are interested in the case where the electric field is switched off. Thus, our non-linear Lagrangian is purely magnetic where we consider $f({\cal F})$ as a functional depending on two real parameters $\alpha$ and $\beta$ given by: 

\begin{eqnarray}
f ({\cal F}) & = & \left( \beta {\cal F}^{- (1+\alpha)} + 1 \right)^{\frac{1}{1+\alpha}},
\end{eqnarray}
where ${\cal F} = \frac{1}{2} B^2$, $\beta {\cal F}^{-(1+\alpha)}$ is dimensionless and for $\beta=0$, we have $f({\cal F})=1$ (i.e. ${\cal L} = - {\cal F}$) which is the usual electrodynamics Lagrangian. \newline

In this purely magnetic case, the energy density and pressure become,
\begin{eqnarray}
\rho  = & {\cal F} \left( \beta {\cal F}^{-(1+\alpha)} + 1 \right)^{\frac{1}{1+\alpha}},  \nonumber \\
p  = & - {\cal F} \left( \beta {\cal F}^{-(1+\alpha)} + 1 \right)^{\frac{1}{1+\alpha}} \nonumber \\
& + \frac{4}{3} {\cal F} \left( \beta {\cal F}^{- (1+\alpha)} + 1 \right)^{-1+\frac{1}{1+\alpha}}.
\end{eqnarray}
\noindent The equation of state satisfied by the above nonlinear electrodynamics Lagrangian density is: 
\begin{eqnarray}
p & = & \frac{1}{3} \rho \left(1 - 4 \frac{\beta}{\rho^{1+ \alpha}} \right)  
\label{eosB}
\end{eqnarray}
\noindent which is obviously the modified Chaplygin gas introduced in \cite{benaoum2002,benaoum2012}, and discussed in detail in the previous section. It is easy to see that when the non-linearity is switched off (i.e., $\beta =0)$, the above equation of state reduces to the Maxwell radiation EoS. \newline

The evolution of the magnetic field is governed by the continuity equation:
\begin{eqnarray}
\dot{\rho} + 3 H \left( \rho + p \right) & = & 0
\end{eqnarray}
By using equation~\eqref{eosB}, a general solution of the above equation in terms of the scale factor is obtained as:
\begin{eqnarray}
\rho & = & \rho_0 \left( \frac{\beta}{\rho_0^{1+ \alpha}} + (1 - \frac{\beta}{\rho_0^{1+ \alpha}}) a^{-4 (1+ \alpha)} \right)^{\frac{1}{1+ \alpha}}  
\label{rhoB}
\end{eqnarray}
where $\rho_0 = \rho (a=1)$ is the present energy density. By comparing equations~\eqref{rho} and \eqref{rhoB},
we find the relation between the non-linearity $\beta$ and the parameter $\gamma_s$ which is given by:
\begin{equation}
\beta =  (1 - \gamma_s) \rho_{de,o}^{1+\alpha} = (1 - \gamma_s) (\Omega_{de,o} \rho_{\text{crit}})^{1+\alpha}
\end{equation}

with $\rho_{\text{crit}} =$ 1.88$\times$10$^{-29}$ 
h$_o^2$ $\cdot$ g $\cdot$ cm$^{-3}$.

\section{Statistical analysis}\label{section3}

In order to study the cosmic evolution of this type of dark energy and compute model predictions for the free parameters, we use the software \textsc{CosmoSIS}\footnote{\url{https://bitbucket.org/joezuntz/cosmosis/wiki/Home}} \cite{cosmosis}. The code provides a Bayesian framework that allows the user to implement different modules within a pipeline that combines a set of observational detections and samplers to evaluate the likelihood function and calculate the best estimates of the model.\newline
\textsc{CosmoSIS} computes the pipeline in discrete steps, performed by independent modules. Among these modules, we include: \textsc{Consistency}, which checks for cohesion and consistency among the defined cosmological parameters and assesses that set of criteria is not under or overdetermined; \textsc{Camb}, a Boltzmann code that calculates the underlying background quantities and computes the linear matter power spectrum \cite{lewis2000,howlett2012}; \textsc{Growth} evaluates the linear growth factor and growth rate within a Universe with a state equation \eqref{eos1}. We run \textsc{Camb} from $z =$ 5000 down to redshift 0 to closely follow the evolution of the components in the domination eras that are affected by the introduction of our model. The 500000 steps in redshift are dynamic to cover evenly the redshift range.\newline

On the other hand, we choose the \textsc{Emcee} sampler\footnote{\url{http://dan.iel.fm/emcee/}} \cite{emcee}, a Monte-Carlo Markov Chain that explores the parameter space with a collection of \textit{walkers}. Each \textit{walker} takes a random path and jumps to the next point after calculating a Metropolis acceptance rate. The final acceptance fraction should be in the range of 0.2-0.5 to reach the maximum posterior distribution \cite{gelman1996}. The total number of samples in the chain comes from the product between the number of walkers and samples. We set 256 samples, 512 walkers, 20 steps before calculating the accuracy rate, and a random start for all \textit{walkers}. The latter condition delays the convergence to the chain; thus, we remove the first 20000 realizations (burn-in stage).\newline

The choice of the hyperparameters of the Bayesian estimator shown above obeys three criteria: i) a high success rate in the convergence of the chain; ii) posterior probability distributions of the physical parameters are unimodal; and iii) probabilities densities are not poorly populated.\newline

The uniform prior distributions for the dark energy and the cosmological parameters imposed in \textsc{CosmoSIS} are presented in Table~\ref{table:priors}. Although the parameter $\alpha$ can take values up to 10, we limit the upper value for its prior distribution to be 0.5 to ensure the classical stability condition discussed in Section~\ref{section2}.
\begin{table}[h!]
\centering
\caption{Uniform prior ranges for the model's free parameters: $\gamma_s$ and $\alpha$ and the cosmological parameters assumed in the analysis.}
\label{table:priors}
\begin{tabular}{|c|c|}\hline 
Parameter   & Prior\\ \hline 
$\gamma_s$   & $[0,1]$\\
$\alpha$   & $[-0.999,0.5]$\\
$\Omega_m$ &  $[0.2,0.4]$\\
h$_{o}$ &  $[0.6,0.8]$\\
$\sigma_8$ &  $[0.5,0.9]$\\\hline
\end{tabular}
\end{table}

The predictions of the model are calculated with precomputed likelihoods available in \textsc{CosmoSIS}:  CMB data with Planck 2018 (TTTEEE + lensing) \cite{planck2020}, primordial nucleosynthesis (or BBN) \cite{beringer2012,cooke2016,pitrou2021}, distances inferred from BAO datasets: 6dF Galaxy Survey -or 6dFGS- \cite{beutler2011,beutler2012}, WiggleZ \cite{kazin2014}, SDSS DR7 (the main galaxy survey) \cite{ross2015}, and luminosity distances from SNIa from Pan-STARRS1 \cite{scolnic2018} -Pan\-theon- and the SH0ES survey \cite{riess2022}. \newline

We run \textsc{CosmoSIS} in the following combined sets of observations:\newline
\noindent\textbf{Set 1}: Planck 2018 (TTTEEE + lensing). \newline
\noindent\textbf{Set 2}: Planck 2018 (TTTEEE + lensing) + BBN + SDSS DR7 (main galaxy survey) + 6dFGS + WiggleZ + Pantheon + SH0ES.\newline

The former setting sets strong constraints for the model at high redshift, whilst the second one acknowledges the cosmological evolution with this form of dark energy at different stages of the Universe. \newline
We perform additional tests with other combined datasets but skip presenting them here because the results are not informative; thus, we cannot set tight constraints to our free parameters. In particular, combined analysis with BAO and SNIa datasets are prior-dominated, and the parameter space is loosely explored in these cases.\newline

The best fits for our model parameters are presented in Table~\ref{table:without_m}. We constrain the value of $\gamma_s$ and $\alpha$ that define the evolution of the dark energy model, three cosmological parameters $\{\Omega_m$,h$_{o}$,$\sigma_8\}$ and $m$, the nuisance parameter that accounts for the SN1a absolute magnitude. Based on the concordance model, we recover the best-fit value for $\Omega_{de,o}$ following the condition $\Omega_{m} + \Omega_{de,o} =$ 1\footnote{We remind the reader that a spatially flat Universe is assumed throughout the document.}.
\begin{table}[h!]
\centering
\caption{Mean value (best-fit) $\pm$ 1$\sigma$ errors of the cosmological parameters of our early dark energy model, obtained with combined analysis of Planck 2018 (TTTEEE+lensing), BBN, SDSS DR7, 6dFGS, WiggleZ, Pantheon, SH0ES data.}
\resizebox{0.48\textwidth}{!}{%
\begin{tabular}{|ccc|}
\hline
 & \textbf{Set 1} & \textbf{Set 2} \\ \hline
$\Omega_m$ & 0.279(0.281)$\pm$0.017 & 0.272(0.271)$\pm$0.017 \\ 
h$_{o}$ & 0.698(0.697)$\pm$0.009 & 0.704(0.702)$\pm$0.009 \\
$\sigma_8$ & 0.805(0.804)$\pm$0.001  & 0.797(0.798)$\pm$ 0.007 \\ 
$\gamma_s$ & 0.508(0.503)$\pm$0.027 & 0.469(0.468)$\pm$ 0.012 \\ 
$\alpha$ & -0.954(-0.957)$\pm$0.021 & -0.942(-0.947)$\pm$0.032 \\ 
$m$ & - & -19.353(-19.352)$\pm$0.023 \\
\hline
$\Omega_{de,o}$ & 0.721(0.719)$\pm$0.017 & 0.728(0.729)$\pm$0.017\\
 \hline
\small{Acceptance rate} & 0.376 & 0.398 \\ \hline
\end{tabular}
}
\label{table:without_m}
\end{table}

\begin{figure*}[h!]
\centering
  \includegraphics[scale=0.71]{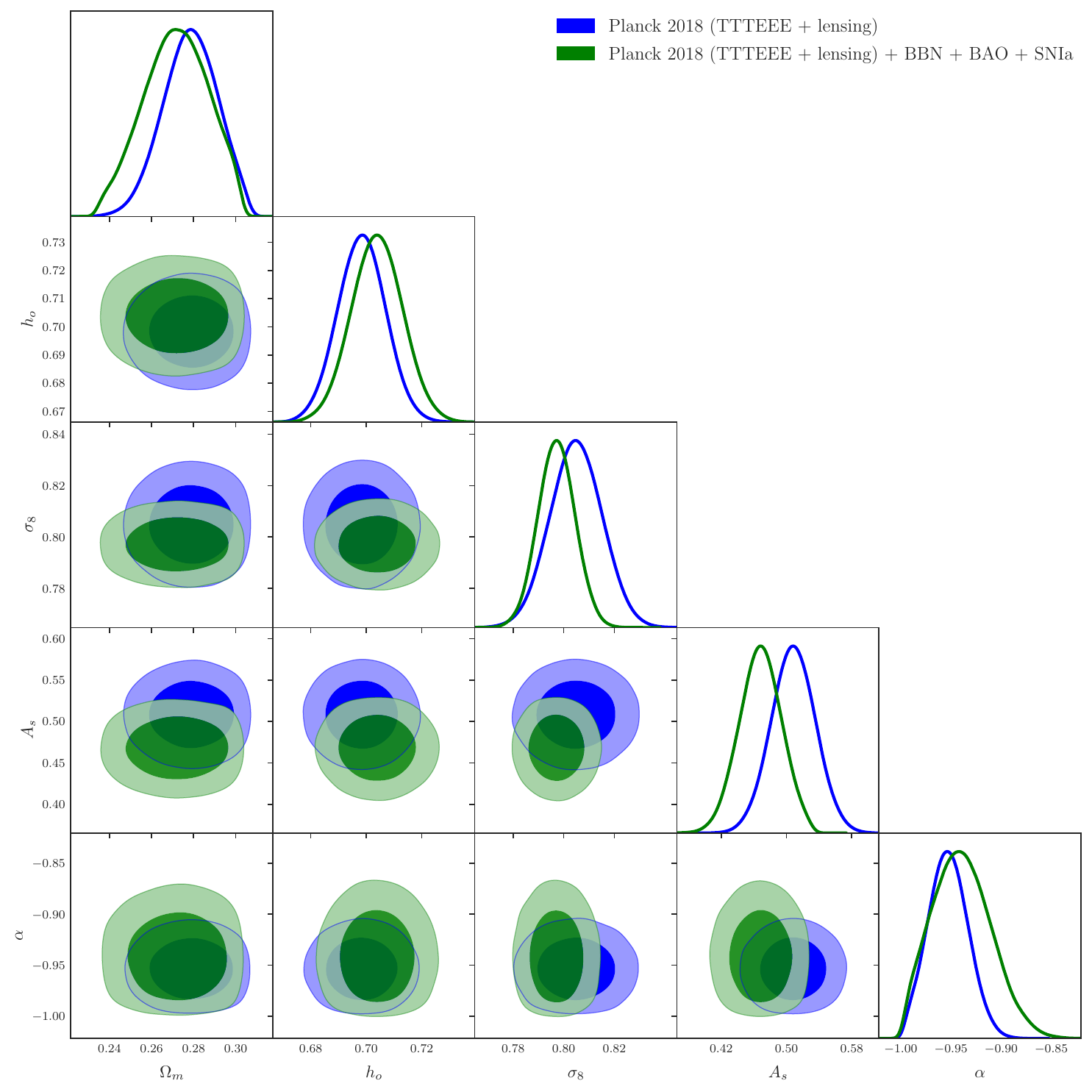} 
\caption{Marginalized posteriors for the proposed early dark energy model and other cosmological parameters after performing a Bayesian analysis with likelihoods from Planck 2018 (TTTEEE + lensing) \cite{planck2020} -blue contours-, and Planck 2018 (TTTEEE + lensing) 
in combination with BBN \cite{beringer2012,cooke2016,pitrou2021}, BAO 6dFGS \cite{beutler2011,beutler2012}, WiggleZ \cite{kazin2014}, SDSS DR7 \cite{ross2015}, and SNIa: Pantheon \cite{scolnic2018} and SH0ES \cite{riess2022}, in green.}
\label{fig:6}
\end{figure*}

The statistical analysis and visualization of the samples are performed with \textsc{GetDist}\footnote{\url{https://getdist.readthedocs.io/en/latest/plots.html}} \cite{lewis2019}. Figure~\ref{fig:6} shows the posterior distribution of the model and cosmological parameters with the combined datasets described above.\newline

\begin{figure}[h!]
\centering
\includegraphics[width=1.1\columnwidth]{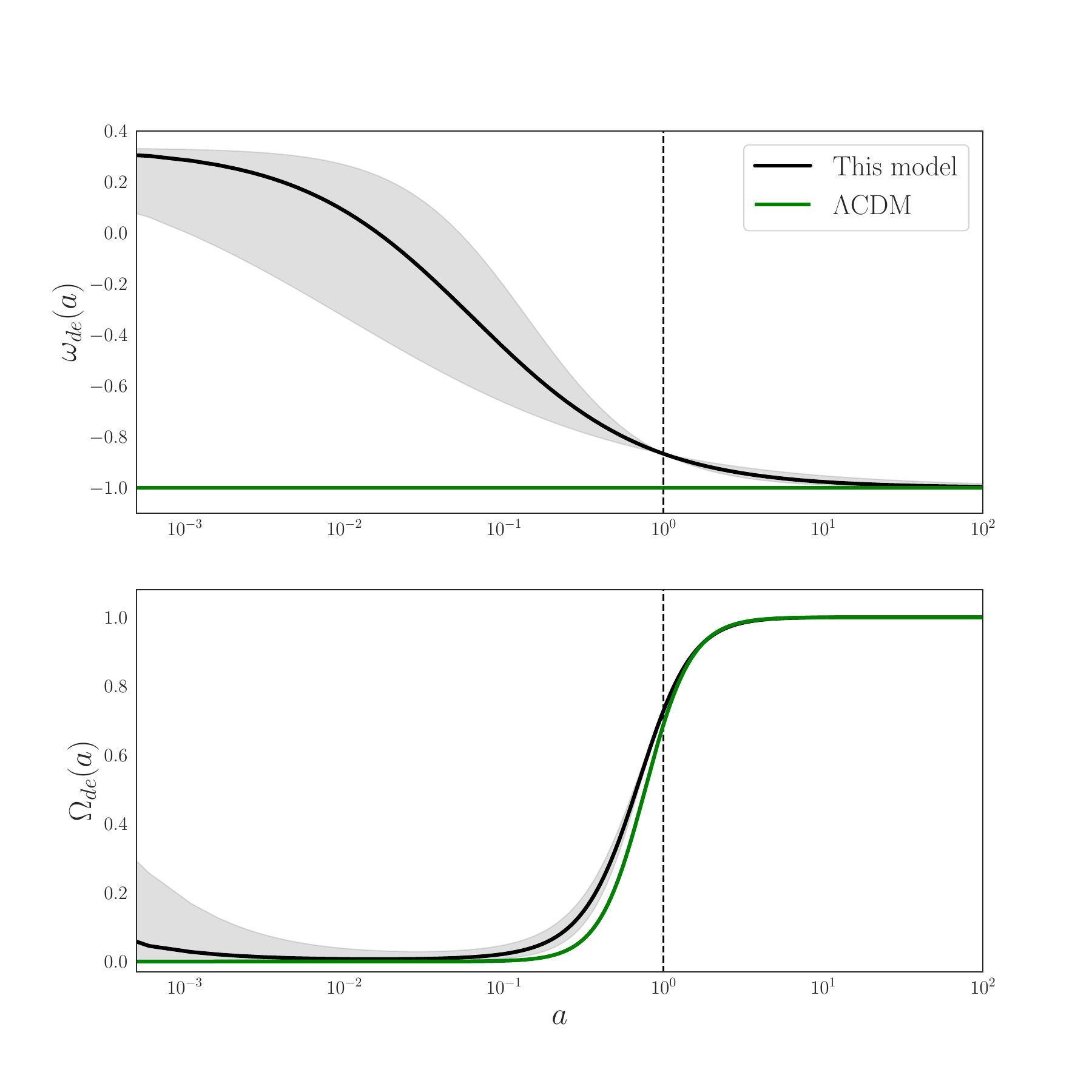}
\caption{Evolution of the equation of state parameter $\omega_{de}$ and energy fraction for our dark energy model $\Omega_{de}$ with the scale factor $a$ in the top and bottom panels, respectively. We have set the values of the best fits presented in Table~\ref{table:without_m} in both plots. The dashed line represents the cut today. The grey shadow region shows error bars for both quantities, based on the errors reported in Table~\ref{table:without_m}, and the green solid lines, the corresponding prediction with the $\Lambda$CDM model with cosmological parameters from the Planck collaboration \cite{planck2020}.}
\label{fig:7}
\end{figure}

Figure~\ref{fig:7} shows the evolution of our model's equation of state $\omega_{de}$ and dark energy density fraction $\Omega_{de}$.  In the upper panel, the equation of state exhibits a transition from the radiation domination epoch ($\omega_{de} \sim \frac{1}{3}$ at ln $a <<$ -3) to the De-Sitter era (ln $a \sim$ 0).\newline
Conversely, the dark energy density fraction as a function of the scale factor $a$ is displayed in the bottom panel of Figure~\ref{fig:7}, according to the expression:
\begin{equation}
\Omega_{de}=\frac{\rho_o}{\rho_{cr}}=\frac{\Omega_{de,o}G (a)}{E^2(a)},
\end{equation}
\noindent where $E(a)$ is defined in equation~\eqref{e_z}, and $G(a)$ in the expression~\eqref{param}. \newline

The bottom panel reveals that today's dark energy density fraction value is $\sim$ 0.7, increasing rapidly in the future (i.e., $a >$ 1). Even more interesting, the energy fraction of the dark energy in this model is always above this quantity associated with $\Lambda$. A natural consequence of this trend is that the matter-dark energy equality occurs earlier, so the structure formation than in the standard model case. In addition, if the amount of dark energy increases at all times with respect to other components of the matter-energy content of the Universe, the position and height of the acoustic peaks would shift with respect to the standard model prediction.\newline

We calculate the numerical value of $\beta = $ 0.52 $\times \rho_{\text{crit}}^{1+\alpha} =$ 0.015 $\pm$ 0.008, the parameter that quantifies the non-linearity of the electrodynamics. Considering the small (and negative) value for $\alpha$ derived from our analysis, the exponent accompanying the factor $\rho_{\text{crit}}$ is significantly small; thus, $\beta$ has a numerically non-zero value but still is a perturbative parameter. This result indicates a departure from Maxwell's classical electrodynamics in the early Universe that could explain the origin of this type of dark energy with a non-negligible contribution during the radiation domination epoch.\newline

Additionally, we compute the value $z_t$ -corresponding to the scale factor $a_t$ defined in equation~\eqref{at}- with our best fits free parameters and find a value for $z_t =$ 0.47. We can read the latter result as an earlier transition to the accelerated expansion epoch than in the $\Lambda$CDM model if the latter would have had dynamics associated to $\Lambda$.\newline

Furthermore, we evaluate $S_8=\sigma_8\sqrt{\Omega_m/0.3}$. Plugging the best fits in our model, we obtain a value of $S_8=$ 0.758 $\pm$ 0.007, fairly consistent with results reported by KiDS-450 and  KiDS-450+2dFLenS \cite{conti2017,joudaki2017,joudaki2018}. However, we find that this quantity, which measures the clustering of the structure, is not the best observable to set tight constraints in this kind of cosmology. Our reports show no tension between the low and high Universe for $\sigma_8$.\newline

Finally, it is important to stress that we do not calculate the matter perturbations due to the dark energy's early contribution to the cosmic plasma. Instead, we derive the parameter $\sigma_8$ as a product of the computations made by \textsc{Camb}. The complete treatment of the density field beyond the background level is outside the scope of this work.

\section{Cosmological tests}\label{section4}

Based on the best fits parameters of the dark energy model presented best-fits presented in Table~\ref{table:without_m} and Figure~\ref{fig:6}, we calculate the age of the Universe, following equation (19) in \cite{boylan2021}:
\begin{equation}\label{to}
t_o =\frac{2}{3}\frac{1}{H_o \sqrt{1 - \Omega_m}}\text{ln}\left(\sqrt{\frac{1}{\Omega_m} - 1} + \sqrt{\frac{1}{\Omega_m}}\right).
\end{equation}
\noindent When $H_o$ is expressed in Gyr$^{-1}$, the age of the Universe in this model is $t_o =$ 13.4 $\pm$ 0.2 Gyr. Instead, the time of the Universe inferred from Planck 2020 \cite{planck2020} is $t_o =$ 13.797 $\pm$ 0.023 Gyr. However, early dark energy models exhibit shorter times (see results from \cite{garcia2021,boylan2021}), leading to an earlier structure formation in the Universe when compared with the $\Lambda$CDM model. This result is consistent with our calculation for $z_t$ in the previous section.\newline

Recently, \cite{jiaqi2023} presented robust estimates of the absolute age of the globular cluster M92. Their analysis shows this cluster's age of 13.80 $\pm$ 0.75 Gyr. Evidently, no structure could have been formed before the Universe itself; hence, their main finding rules out certain cosmological models in which the age of the Universe is outside the age range for M92. Our model passes the test, lying in the lower limit defined by \cite{jiaqi2023}.\newline

There is an additional proxy that we submit our model, assuming the lookback time ($t_{\text{lb}}$) as a function of the scale factor ($a$), first explored in \cite{boylan2021}:
\begin{align}\label{tlb}
t_{\text{lb}}(a) & =\frac{2}{3} \frac{1}{H_o \sqrt{1 - \Omega_m}} \text{arcsinh}\left(\sqrt{\frac{1}{\Omega_m} - 1}\right) \nonumber\\
& -\frac{2}{3} \frac{1}{H_o \sqrt{1 - \Omega_m}}\text{arcsinh}\left(\sqrt{\frac{1}{\Omega_m} - 1}a^{3/2}\right).
\end{align}
\noindent Using the expression \eqref{tlb}, we calculate the time of the completion of the Epoch of Reionization (EoR), under the assumption that 6 $< z_{\text{EoR}} <$ 10 \cite{madau2014,stark2016,greig2017,garcia2017}. The lookback time predicted by our model to the duration of the cosmic Reionization is 12.37 $>t_{\text{lb, EoR}}/\text{Gyr}>$ 12.93. \newline

\begin{figure}[h!]
\centering
  \includegraphics[width=1.1\columnwidth]{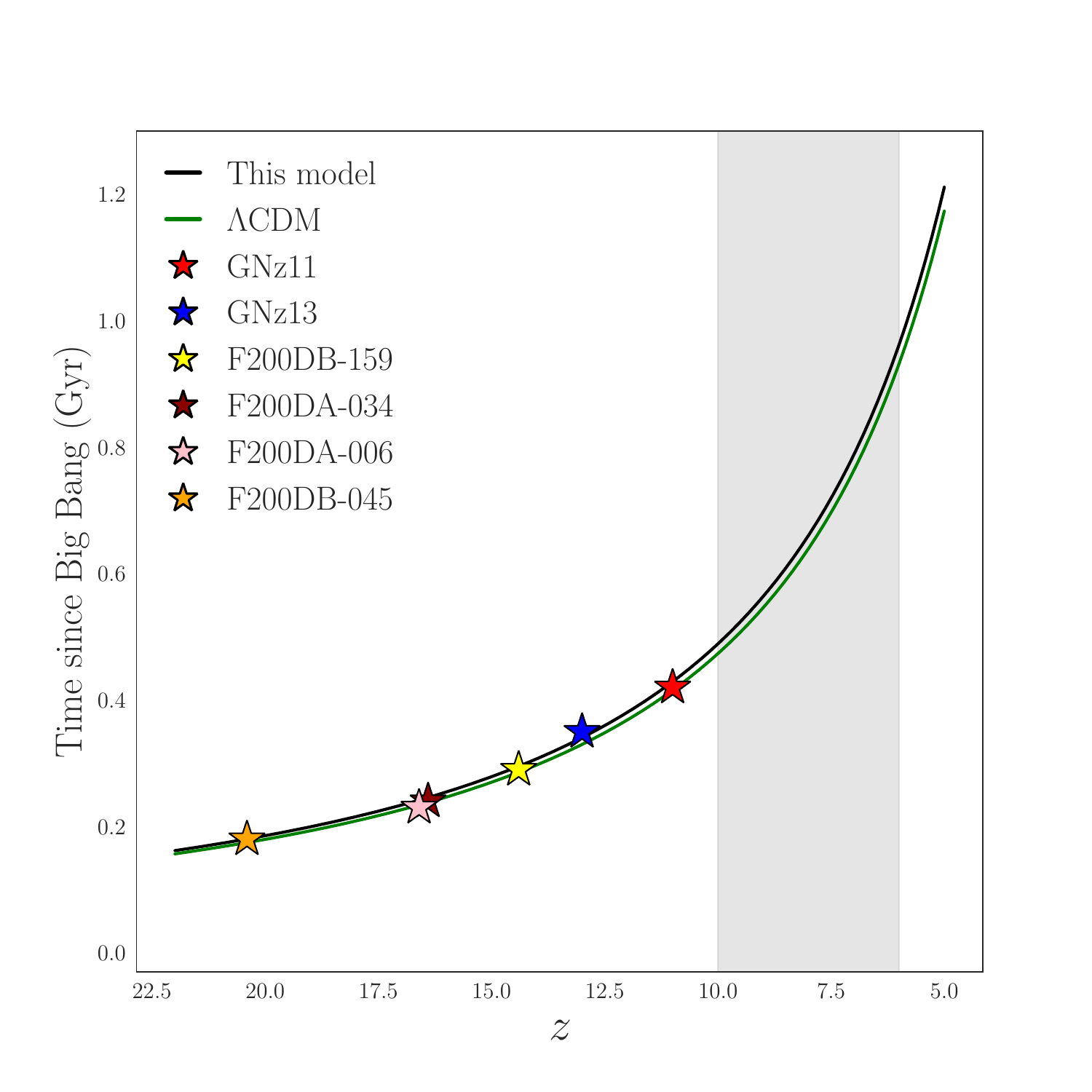}
\caption{Lookback time calculated with equation~\eqref{tlb} as a function of $z$, for our model best-fit parameters (see Table~\ref{table:without_m}) in black and the standard cosmological model with Planck cosmological parameters \cite{planck2020} in green. We compare our theoretical predictions to the EoR \cite{madau2014,stark2016,greig2017,garcia2017} in grey shadow region, GNz-11 \cite{jiang2021} and the highest redshift galaxy candidates presented by GLASS \cite{yan2022} in the red stars.}
\label{fig:lookback}
\end{figure}

Moreover, we present the lookback time evolution with $z$ for our model's best parameters in Figure~\ref{fig:lookback} and the same function for the $\Lambda$CDM model. We include a conservative duration for the EoR: 6 $< z_{\text{EoR}} <$ 10, the highest redshift galaxy detected by HST: GNz-11 \cite{jiang2021} and the highest $z$ ga\-la\-xy candidates observed with the JWST, by the GLASS collaboration \cite{yan2022}. We remind the reader that the calculation of the lookback time for these galaxies is model-dependant because the observable of these objects is the redshift; therefore, we need to assume a cosmology to compute the time that their light has been traveling before reaching the telescope. In this case, we consider the set of parameters by \cite{planck2020} as the fiducial cosmology for these candidates at high redshift.

Lastly, we determine the CMB shift parameter R$_{\text{CMB}}$, which measures the shift of the first acoustic peak in the power spectrum of the CMB anisotropies. We numerically calculate this observable using the dimensionless Hubble parameter $E(z)$, equation \eqref{e_z}, which can be expressed as:
\begin{equation}
    R_{\text{CMB}} = \Omega_m^{1/2} \int_0^{1089} \frac{dz}{E(z)}.
\end{equation}
\noindent With the best fit parameters reported in Section~\ref{section3}, this quantity has a value of R$_{\text{CMB}} =$ 1.61$^{+0.23}_{-0.04}$. This estimate is lower than the value inferred from the different Planck campaigns and is consistent with a dark energy model that produces an early structure evolution. Therefore, it shifts slightly between the first and second acoustic peaks.\newline

We close this section with an important remark: introducing a form of dark energy originated by a departure from Maxwell's electrodynamics causes matter structure to form more promptly than in the standard cosmological model. This variation is mainly regulated by a non-zero $\gamma_s$ parameter (or equivalently, $\beta \neq$ 0). Nonetheless, if the $\gamma_s$ parameter is switched off, $\gamma_s =$ 0, then the $\Lambda$CDM cosmology and the classical Maxwell electrodynamics are fully recovered.
 
\section{Conclusions}\label{section5}
We propose an early dark energy model that mimics radiation at the early stages of the Universe and exhibits an accelerated expansion in the late evolution of the Universe. We show that such an effective parametrization can be nicely modeled within the non-linear electrodynamics framework where the non-linearly plays important during the evolution of the Universe. The model assumes three free parameters $\{\gamma_s$,$\alpha$,$\Omega_{de,o}\}$, and a parameter $\beta$ that relates to the non-linearity of the electromagnetic tensor, heavily dependent on the value of $\gamma_s$.\newline

We find the best-fit parameters for this model, combining different sets of cosmological observations: CMB, BBN, BAO, and SNIa distances, and their likelihoods in the modular code \textsc{CosmoSIS}. The best values for the free parameters are: $\gamma_s =$ 0.468 $\pm$ 0.026, $\alpha =$ -0.947 $\pm$ 0.032, and $\Omega_{de,o} =$ 0.729 $\pm$ 0.017 (derived parameter from the MCMC best fit for $\Omega_{m}$ and the assumption of the Concordance model). The departure from the Maxwell electrodynamics is quantified with the derived parameter $\beta=$ 0.015 $\pm$ 0.008. The non-null value of $\beta$ indicates that contributions from non-linear electrodynamics in the early Universe could originate this form of dark energy. In addition, we set constraints for $\Omega_{m}=$ 0.271 $\pm$ 0.017, h$_o =$ 0.702 $\pm$ 0.009, and $\sigma_8 =$ 0.798 $\pm$ 0.007. \newline

Based on the Bayesian analysis presented in this document, we demonstrate that the Universe experiences a faster expansion rate during the radiation domination epoch, induced by additional degrees of freedom in the Hubble parameter. Our Universe subjected to this type of dark energy enters the matter-domination epoch earlier than predicted by the standard model. Thus, our model foresees an earlier structure formation. This result is consistent with findings from recent works \cite{smith2021,klypin2021,boylan2021} that explore the possibility that early dark energy models explain the current Hubble tension, which has reached the 5.7$\sigma$ discrepancy between Planck 2018 + $\Lambda$CDM and late time measurements of H$_o$ \cite{review2022,cocoa2022}. As presented in this document, a scenario of the Universe with an early dark energy contribution alleviates the tension from comparing early and late Universe datasets \cite{gomez2021}. The value for H$_o$ reported here sits between estimates from Planck 2018 and SH0ES. The latter result is particularly encouraging to pursue future studies with this form of dark energy.\newline

Also, we offer an estimate of the $S_8$ parameter, frequently associated with a tight correlation between $\sigma_8$ and $\Omega_m$. The prediction from our model is $S_8=$ 0.758 $\pm$ 0.007, consistent with the value reported by KiDS and the 2dFLenS. However, we stress it is outside of the scope of the paper to study the evolution of the matter perturbations within this model.\newline

Finally, our model is a compelling formalism to narrow the current cosmological tensions between different datasets. Future generations of missions, such as the Nancy Grace Roman Space Telescope (WFIRST) and EUCLID, will shed light on the nature of dark energy and its dynamics in the early Universe.

\begin{acknowledgements}
H.B.Benaoum gratefully acknowledges the financial support from the University of Sharjah. L.A. Garc\'ia thanks the valuable contribution from Joe Zuntz for his insightful remarks to implement the software \textsc{CosmoSIS} in private communications. L. Cas\-ta\-\~ne\-da was supported by Patrimonio Aut\'onomo - Fondo Nacional de Financiamiento para la Ciencia, la Tecnolog\'ia y la Innovaci\'on Francisco Jos\'e de Caldas (MINCIENCIAS - COLOMBIA) Grant No.\newline
110685269447 RC-80740-465-2020, projects 69723.
\end{acknowledgements}

\end{document}